\documentclass[english,twocolumn,amsmath,amssymb,aps,pra,showpacs]{revtex4-1}
\usepackage[T1]{fontenc}
\usepackage[latin9]{inputenc}
\usepackage{amsmath}
\usepackage{amssymb}
\usepackage{graphicx}
\usepackage{esint}
\usepackage{babel}
\begin{document}

\title{Ultra-cold Collisions between Spin-Orbit-Coupled Dipoles: General
Formalism and Universality}

\author{Jia Wang}

\affiliation{Centre for Quantum and Optical Science, Swinburne University of Technology,
Melbourne 3122, Australia}

\author{Christiaan R. Hougaard}

\affiliation{Centre for Quantum and Optical Science, Swinburne University of Technology,
Melbourne 3122, Australia}

\author{Brendan C. Mulkerin}

\affiliation{Centre for Quantum and Optical Science, Swinburne University of Technology,
Melbourne 3122, Australia}

\author{Xia-Ji Liu}

\affiliation{Centre for Quantum and Optical Science, Swinburne University of Technology,
Melbourne 3122, Australia}

\date{\today}
\begin{abstract}
A theoretical study of the low-energy scattering properties of two
aligned identical bosonic and fermionic dipoles in the presence of
isotropic spin-orbit coupling (SOC) is presented. A general treatment
of particles with arbitrary (pseudo-) spin is given in the framework
of multi-channel scattering. At ultracold temperatures and away from
shape resonances or closed-channel dominated resonances, the cross-section
can be well described within the Born approximation to within corrections
due to the $s$-wave scattering. We compare our findings with numerical
calculations and find excellent agreement.
\end{abstract}
\maketitle

\section{Introduction}

Many novel behaviors and phases in quantum few- and many-body systems
can be understood from the competition between kinetic and interaction
energies. The extraordinary degree of control in systems of ultracold
quantum gases, therefore, provides versatile platforms to study these
quantum phenomena \cite{RMP-bose-1,RMP-bose-2,Bloch2008RMP,RMP-fermi}.
For instance, short-range interactions (or more precisely, $s$-wave
scattering lengths) between atoms can be tuned to virtually arbitrary
values via magnetic Feshbach resonances \cite{ChinReview2010}, allowing
us to access the unitary regime \cite{OHaraThomas2002Science,RakotynNPhys2013,HuMG2016PRL}
and study Efimov physics \cite{efimov,Berninger2011PRL,JiaWang2012PRL}.
On the other hand, long-range interactions, especially the anisotropic
dipole-dipole interactions, can significantly change the excitation
spectrum \cite{Santos2003PRL} and the stability diagrams of Bose-Einstein
Condensation (BEC) \cite{Santos2000PRL,Griesmaier2005PRL,Koch2008NatPhys},
which has attracted intense interest in gases of ultracold heteronuclear
ground state molecules \cite{Ni2008Science,RvachovKetterle2017PRL},
and magnetic dipolar atoms such as $^{52}$Cr \cite{Griesmaier2005PRL},
$^{164}$Dy \cite{Lu2010PRL,Lu2011PRL}, and $^{168}$Er \cite{Aikawa2012PRL}.
More recently, manipulating kinetic energies and corresponding dispersion
relationship has been realized via the innovative synthetic spin-orbit
coupling (SOC) technique, i.e. coupling a particle's canonical momentum
with its (pseudo-) spin degrees of freedom \cite{Zhangjing2014ReviewSOC,Zhaihui2015ReviewSOC}.
The realization of SOC provides an important ingredient for studying
the fractional quantum Hall effect, topological insulators \cite{Hasan2010RMP,Qi2011RMP},
and has been a fundamental advancement in ultracold quantum gases
in recent years \cite{Dalibard2011RMP,Goldman2014RPP}.

There are currently several experimental techniques to realize SOC
in cold-atom systems, such as lattice shaking \cite{Struck2012PRL}
and Raman coupling \cite{Lin2011Nature}. In particular, the Raman
laser scheme has been applied to achieve one-dimensional SOC (an equal
mixture of Rashba and Dresselhaus spin-orbit coupling) \cite{Lin2011Nature,Cheuk2012PRL,Wang2012PRL,Zhang2012PRL,Chunlei2013PRA,Khamehchi2017PRL}
and two-dimensional SOC \cite{Huang2016NatPhys,Meng2016PRL,Wu2016Science}
in ultracold gases of alkali atoms. However, Raman coupling for alkali
atoms usually also comes along with atomic heating due to spontaneous
emission. This heating leads to the loss of quantum degeneracy and
trap population, which is a major challenge to study many-body quantum
phenomena that manifest at a lower temperature and longer timescales.
For atoms with higher ground state orbital angular momentum, spontaneous
emission can be eliminated while still producing large Raman coupling
\cite{Cui2013PRL}, making the open-shell lanthanide atoms Dy and
Er suitable candidates. These elements also possess large magnetic
dipole moment, allowing studies of the interplay between SOC and long-range
dipole-dipole interactions that do not exist in alkali atomic gases,
but requiring a more sophisticated theoretical model. Experimentally,
SOC has been recently achieved by Ref. \cite{Burdick2016PRX} in $^{161}$Dy,
which allowed for the realization of a long-lived SOC degenerate dipolar
Fermi gas. The bosonic system of SOC dipolar gas has also been theoretically
investigated previously in BEC of $^{52}$Cr \cite{Deng2012PRL}.

The existence of \textit{long-range} dipole-dipole interactions in
these systems is expected to give an interesting interplay with the
SOC, leading to intriguing new quantum phases. Previous theoretical
studies on the interplay between \textit{short-range} two-body interactions
and SOC in Fermi gases have explored novel superfluid states in the
BEC-BCS crossover \cite{Gong2011PRL,HuHui2011PRL,Yu2011PRL,Vyasanakere2011PRB,Han2012PRA}.
For a BEC with SOC, new quantum phases, such as a stripes phase, have
been predicted for a certain range of the Raman coupling strengths
determined by the inter- and intraspecies scattering lengths \cite{WangZhai2010PRL,Ho2011PRL,ZhangZhang2012PRL,LiYun2013PRL}.
The anisotropic and long-range dipole-dipole interaction can be regarded
as an additional degree of freedom, which might lead to new physics,
but also brings new challenges in theoretical studies. To construct
a concrete theoretical model for SOC dipolar quantum gases, the low-energy
scattering between two dipoles in the presence of SOC needs to be
understood first, which is the main topic of our study here.

Our theoretical formalism is inspired by several previous studies
on ultracold collisions between two non-dipole particles in the presence
of the three-dimensional (3D) isotropic SOC (which is a 3D analog
of Rashba SOC) \cite{Cui2012PRA,Zhang2012PRA,Duan2013PRA,WangSuJu2015PRA,Guan2016PRA,Guan2017PRA}.
While the 3D isotropic SOC has not yet been realized experimentally
in cold-atom systems, proposals have been made that are based on adding
more laser fields. Recently, the realization of 2D isotropic SOC using
this scheme has been reported \cite{Sun2017}. The laser scheme to
realize 3D isotropic SOC is ideally suited for lanthanide atoms, where
there is less atomic heating. On the other hand, the 3D isotropic
SOC is more closely related to the cases in condensed-matter physics
due to the high symmetry \cite{Wan2011PRB,Burkov2011PRL}. This symmetry
also allows for a fully analytical treatment of low-energy scattering
in the presence of SOC.

The low-energy scattering is strongly affected by the asymptotic behavior
of atoms at large distances. In our system, the SOC persists even
for atoms with infinite separation, which changes the threshold energies,
and modifies the dispersion relation \cite{Dalibard2011RMP}; on the
other hand, a dipole-dipole interaction also dominates potential energies
at large distances. The competition between SOC and dipole-dipole
interaction, therefore, gives rise to special threshold behaviors.
The details of our formalism to study this problem is outlined below
in Sec. II. An analysis by applying the first-order Born approximation
is given in Sec. III. A comparison with numerical results for spin-$1/2$
dipolar fermions is given in Sec. IV. A summary of our study is given
in Sec. V.

\section{Formalism}

In our model, each dipole is treated as a point particle with mass
$m$. The interaction potential $V(\vec{r})$ between two dipoles
aligned to the $z$-axis and seperated by a large distance $\vec{r}$
is therefore given by $V(\vec{r})\rightarrow V_{d}\left(\vec{r}\right)=d^{2}\left(1-3\cos^{2}\theta\right)/r^{3}$.
Here $\theta$ is the polar angle of $\vec{r}$ in spherical coordinates,
and $d=\mu_{m}\sqrt{\mu_{0}/4\pi}$ denotes the dipole moment, where
$\mu_{m}$ is the magnetic dipole moment and $\mu_{0}$ is the vacuum
permeability. The characteristic length scale of the dipole potential
$V_{d}(\vec{r})$ is given by the dipole length $D=\mu d^{2}/\hbar^{2}$
, where $\mu=m/2$ is the two-body reduced mass. Correspondingly,
a natural energy scale can be defined by the dipole energy $E_{D}=\hbar^{2}/\mu D^{2}$.
To mimic the experimentally available control of short-range interactions
by using methods such as Feshbach resonances, we model the short-range
potential by a simplistic hard-wall potential, i.e. $V(\vec{r})=V_{d}(\vec{r})$
for $r\ge r_{c}$, and $V(\vec{r})=\infty$ for $r<r_{c}$. This specific
chosen form of the short-range potential, however, does not limit
the generality of our study of ultracold collisions especially near
potential resonances, which will be elaborated on later. 

Similar to Refs. \cite{Cui2012PRA,Zhang2012PRA,Duan2013PRA,WangSuJu2015PRA,Guan2016PRA,Guan2017PRA},
we focus on the scattering in the center-of-mass frame. With the presence
of 3D isotropic SOC, the Hamiltonian in relative coordinates is given
by,
\begin{equation}
H=\frac{\vec{p}^{2}}{2\mu}+\frac{k_{{\rm so}}}{2\mu}\vec{p}\cdot(\vec{s}_{1}-\vec{s}_{2})+V(\vec{r}),\label{eq:Hamiltonian}
\end{equation}
where $\vec{s_{1}}$ and $\vec{s_{2}}$ are the spin operators for
atom 1 and atom 2, $\vec{p}$ is the relative momentum operator, and
$k_{{\rm so}}$ is the strength of SOC in the units of inverse length.
The energy scale for SOC can therefore be defined by the recoil energy
$E_{r}=\hbar^{2}k_{{\rm so}}^{2}/2m$.

\begin{widetext}Following the same spirit as Refs. \cite{Duan2013PRA,WangSuJu2015PRA,Guan2016PRA,Guan2017PRA},
we solve the relative Schrödinger equation formally as a multichannel
problem, i.e. using channel functions (basis) of $\Omega$, all degrees
of freedom except for $r$, to expand the $\tau$'th independent solution
as,
\begin{equation}
\Psi_{\tau}(\vec{r})=\sum_{\nu}\Phi_{\nu}(\Omega)\frac{F_{\nu\tau}(r)}{r}.
\end{equation}
The channel functions adopted here are the tensor spherical harmonics
that are simultaneous eigenstates of $\{\vec{j}^{2},j_{z},\vec{\ell}^{2},\vec{s}^{2}\}$
whose eigenvalues are collectively represented by $\nu$. Here, $\vec{\ell}$
is the (relative) orbital angular momentum operator, $\vec{s}=\vec{s}_{1}+\vec{s}_{2}$
is the total spin operator, $\vec{j}$ is the total angular momentum,
and $j_{z}$ is the projection to the $z$-axis in the laboratory
frame. The tensor spherical harmonics are defined as,
\begin{equation}
\Phi_{\nu}\left(\Omega\right)\equiv\langle\theta,\phi|(\ell s)jm_{j}\rangle=i^{\ell}\sum_{m_{\ell},m_{s}}C_{\ell m_{\ell};sm_{s}}^{jm_{j}}Y_{\ell m_{\ell}}\left(\theta,\phi\right)\chi\left(s,m_{s}\right),\label{eq:sphericalhamornics}
\end{equation}
where $C_{\ell m_{\ell};sm_{s}}^{jm_{j}}$ are the Clebsch-Gordan
coefficients, $Y_{\ell m_{\ell}}\left(\theta,\phi\right)$ are the
usual spherical harmonics, and $\chi\left(s,m_{s}\right)$ denote
the spin states. The $i^{\ell}$ phase term is introduced to make
the matrix elements of the Hamiltonian all real, which will be convenient
for carrying out numerical propogation later. The matrix elements
for the first two terms in Eq. (\ref{eq:Hamiltonian}) (except for
an additional phase associated with the $i^{\ell}$ term) have been
derived previously in Refs. \cite{Duan2013PRA,WangSuJu2015PRA}:
\begin{equation}
\left\langle (\ell',s')j'm_{j}'\right|\frac{\vec{p}^{2}}{2\mu}\left|(\ell,s)jm_{j}\right\rangle \equiv\frac{\hbar^{2}}{2\mu}\left(-I_{\nu'\nu}\frac{d^{2}}{dr^{2}}+B_{\nu'\nu}^{(2)}\right)=\left(-\frac{\hbar^{2}}{2\mu}\frac{d^{2}}{dr^{2}}+\frac{\hbar^{2}\ell(\ell+1)}{2\mu r^{2}}\right)\delta_{jj'}\delta_{m_{j}m_{j}'}\delta_{\ell\ell'}\delta_{ss'},
\end{equation}
and
\begin{align}
 & \left\langle (\ell',s')j'm_{j}'\right|\frac{k_{{\rm so}}}{2\mu}\vec{p}\cdot(\vec{s_{1}}-\vec{s_{2}})\left|(\ell,s)jm_{j}\right\rangle \equiv\frac{\hbar^{2}}{2\mu}\left(A_{\nu'\nu}\frac{d}{dr}+B_{\nu'\nu}^{(1)}\frac{1}{r}\right)\nonumber \\
 & \ =\delta_{jj'}\delta_{m_{j}m_{j}'}\left\{ \begin{array}{ccc}
s' & 1 & s\\
\ell & j & \ell'
\end{array}\right\} (-1)^{j+\ell+s'+s_{1}+s_{2}}\nonumber \\
 & \ \ \ \times\left[-(-1)^{s}\sqrt{s_{1}(s_{1}+1)(2s_{1}+1)}\left\{ \begin{array}{ccc}
s_{1} & s_{2} & s\\
s' & 1 & s_{1}
\end{array}\right\} +(-1)^{s'}\sqrt{s_{2}(s_{2}+1)(2s_{2}+1)}\left\{ \begin{array}{ccc}
s_{2} & s_{1} & s\\
s' & 1 & s_{2}'
\end{array}\right\} \right]\nonumber \\
 & \ \ \ \times\left(-\frac{\hbar^{2}k_{{\rm so}}}{2\mu}\right)\left[\left(\frac{d}{dr}-\frac{\ell+1}{r}\right)\sqrt{\ell+1}\delta_{\ell',\ell+1}+\left(\frac{d}{dr}+\frac{\ell}{r}\right)\sqrt{\ell}\delta_{\ell',\ell-1}\right],
\end{align}
which are all real. Here, the curly bracket denotes the 6j symbol.
Since the isotropic SOC preserves total angular momentum, different
$j$'s are not coupled by these two terms. However, the anisotropic
dipole-dipole interaction will couple different $j$'s, and only $m_{j}$
is still a good quantum number due to the azimutual symmetry. The
matrix elements for dipole-dipole interaction are then given by, 
\begin{equation}
\left\langle (\ell',s')j'm_{j}'\right|V_{d}(\vec{r})\left|(\ell,s)jm_{j}\right\rangle \equiv\frac{\hbar^{2}}{2\mu}B_{\nu'\nu}^{(3)}\frac{1}{r^{3}}=-i^{\ell-\ell'}\frac{2d^{2}}{r^{3}}(-1)^{\ell'-\ell+s+j}\delta_{s's}\Pi_{\ell j\ell'}C_{jm_{j};20}^{j'm_{j}'}\left\{ \begin{array}{ccc}
\ell & s & j\\
j' & 2 & \ell'
\end{array}\right\} \left(\begin{array}{ccc}
\ell' & 2 & \ell\\
0 & 0 & 0
\end{array}\right),\label{eq:Vdmatelement}
\end{equation}
where $\Pi_{\ell j\ell'}=\sqrt{2\ell+1}\sqrt{2j+1}\sqrt{2\ell'+1}$.
These matrix elements are also real despite the $i^{\ell'-\ell}$
factor, since the 3-j symbol at the end of Eq. (\ref{eq:Vdmatelement})
ensure that $\ell'-\ell=0,\pm2$. In addition, the Clebsh-Gordan coefficient
$C_{jm_{j};20}^{j'm_{j}'}$ shows that $m_{j}$ is a good quantum
number and only channels with $|j-j'|\le2$ can be coupled (specially,
if $m_{j}=0$, only channels with $|j-j'|=0,2$ are coupled.) In principle,
one needs to include channels with all possible angular momenta for
an exact calculation, however, only a finite number of channels, $j\le j_{{\rm max}}$,
are needed in practice to obtain a converged result for the scattering
cross-sections, where $j_{{\rm max}}$ is sufficiently large (about
$40$ for our chosen parameters) \cite{convergence}. \end{widetext}

In terms of these matrix elements, the Schrödinger equation in matrix
form can then be written as, 
\begin{equation}
\left(-\underline{I}\frac{d^{2}}{dr^{2}}+\underline{A}\frac{d}{dr}+\underline{B(r)}-k^{2}\underline{I}\right)\underline{F(r)}=0,\label{eq:radialequation}
\end{equation}
where the underlined variables, $\underline{M}$, denote matrices
with matrix elements $M_{\nu'\nu}$, $\underline{I}$ is the identity
matrix, $E=\hbar^{2}k^{2}/2\mu$ is the incident energy and $\underline{B(r)}=\underline{B^{(1)}}/r+\underline{B^{(2)}}/r^{2}+\underline{B^{(3)}}/r^{3}$.
The logarithmic derivative matrix $\underline{\mathcal{L}}=\underline{F}'\underline{F}^{-1}$
can be obtained by propogating from $r_{c}$ to a sufficiently large
distance $r_{{\rm max}}$ (about $10^{4}D$ for our chosen parameters).
The details of the propagation method is elaborated in Appendix \ref{sec:Propagation-Method}.
The $K$-matrix and $S$-matrix are therefore given by, 
\begin{equation}
\underline{\mathcal{K}}=(\mathcal{\underline{L}}\underline{g}-\underline{g'})^{-1}(\mathcal{\underline{L}}\underline{f}-\underline{f'}),\label{eq:Kmatrix}
\end{equation}
and

\begin{equation}
\underline{\mathcal{S}}=(\underline{I}+i\underline{\mathcal{K}})(\underline{I}-i\underline{\mathcal{K}})^{-1},
\end{equation}
respectively, where $\underline{f}$ and $\underline{g}$ are the
regular and irregular solutions in matrix form. 

The regular solutions $\underline{f}$ are obtained by projecting
the plane-wave solutions onto the tensor spherical harmonics, Eq.
(\ref{eq:sphericalhamornics}). The plane-wave solution with scattering
energy $E=\hbar^{2}k^{2}/2\mu$ can be written as
\begin{multline}
\left\langle \vec{r}\right|{\xi,\zeta;+\hat{k}}\rangle=\sqrt{\frac{1}{2+2\delta_{\xi\zeta}}}\times\\
\left[{e^{i\vec{k}_{\xi\zeta}\cdot\vec{r}}\left|{\xi,+\hat{k}}\right>\left|{\zeta,-\hat{k}}\right>+(-)^{p_{b}}e^{-i\vec{k}_{\xi\zeta}\cdot\vec{r}}\left|{\zeta,-\hat{k}}\right>\left|{\xi,\hat{k}}\right>}\right],
\end{multline}
where $p_{b}$ equals $0$ $(1)$ for identical bosons (fermions)
and $\left|{\xi,\hat{n}}\right>$ is a single-particle state with
the projection of spin along the quantization axis $\hat{n}$ being
$\hbar\xi$. In the presence of 3D isotropic SOC, the particle can
be well described by its helicity state, where the quanization axis
is along the direction of its canonical momentum. Here $\vec{k}_{\xi\zeta}$
is the canonical momenta with direction $\vec{k}$ and magnitude $k_{\xi\zeta}=\sqrt{k^{2}+\kappa_{\xi\zeta}^{2}}-\kappa_{\xi\zeta}$
, where $\kappa_{\xi\zeta}=\left({\xi+\zeta}\right)k_{{\rm so}}/2$.
The expansion gives the matrix elements of $\underline{f}$ as $f_{\nu\tau}=u_{\nu\tau}k_{\tau}rj_{l}(k_{\tau}r)/\sqrt{N_{\tau}}$,
where $N_{\tau}=\pi\hbar^{2}\sqrt{k^{2}+\kappa_{\xi\zeta}^{2}}/2\mu$
is the normalization constant chosen to ensure flux density conservation,
and
\begin{equation}
u_{\nu\tau}=\sqrt{\frac{2\ell+1}{2j+1}}C_{\ell,0;s,\xi-\zeta}^{j,\xi-\zeta}C_{s_{1},\xi;s_{2},-\zeta}^{s,\xi-\zeta}\frac{1+(-)^{s_{1}+s_{2}-s+\ell+p_{b}}}{\sqrt{2+2\delta_{\xi\zeta}}}.\label{eq:utransformation}
\end{equation}
Correspondingly, the matrix elements of the irregular solutions $\underline{g}$
are given by $g_{\nu\tau}=u_{\nu\tau}k_{\tau}rn_{l}(k_{\tau}r)/\sqrt{N_{\tau}}$,
where $j_{l}$ and $n_{l}$ are the regular and irregular spherical
Bessel functions respectively. Hereafter, unless specified otherwise,
we use $\nu=\{j,m_{j},\ell,s\}$ to collectively represent the quantum
numbers of the channel function in Eq. (\ref{eq:sphericalhamornics})
and $\tau=\{j,m_{j},\xi,\zeta\}$ to represent the partial wave of
a particular helicity state. Therefore, Eq. (\ref{eq:utransformation})
can also be regarded as a unitary tranformation between the helicity
basis denoted by $\{\xi,\zeta\}$ and the spin singlet/triplet basis
in the absence of SOC indicated by quantum number $\{\ell,s\}$ for
a particular partial wave of $\{j,m_{j}\}$. The explicit values of
$u_{\nu\tau}$ for spin-$1/2$ fermions and spin-$1$ bosons has been
previously obtained for $j=0$ in Ref. \cite{Duan2013PRA} and Ref.
\cite{WangSuJu2015PRA} respectively, which can be used to verify
our Eq. (\ref{eq:utransformation}) (after carefully taking care of
the $i^{\ell}$ factor). The form of regular and irregular solutions
guarantee the $K$-matrix to be real and symmetric (and hence the
$S$-matrix to be unitary), where the proof is given in Appendix \ref{sec:Symmetry-of--matrix}.

Finally, the cross-section from one partial wave $\{j',m_{j}'\}$
of helicity states $\{\xi',\zeta'\}$ to another partial wave $\{j,m_{j}\}$
of another helicity channel $\{\xi,\zeta\}$ is given by,
\begin{equation}
\sigma_{\tau'\tau}=\frac{2\pi}{k_{\tau'}^{2}}|\mathcal{S}_{\tau'\tau}-\delta_{\tau'\tau}|^{2}.
\end{equation}

\section{First-order Born Approximation}

One of the most important observables in ultracold collisions is the
threshold law behavior determined by the competition between SOC,
short-, and long-range interactions. In this work, we are mostly interested
in magnetic dipolar atoms whose dipole lengths is about $1\sim10$
nm for different species. Therefore, we are focusing on the case $k_{{\rm so}}D<1$,
as the SOC strength is reasonably estimated to be $k_{{\rm so}}\approx1\sim10$
${\rm \mu m}^{-1}$. We remark that the parameter regime of $k_{{\rm so}}D>1$
might be achieved in systems of heteronuclear molecules or Rydberg
atoms, which is, however, beyond the scope of this paper. 

The threshold behaviors for dipolar scattering without SOC have been
discussed in Refs. \cite{Bohn2009NJP,WangYujun2012PRA,Bohn2014PRA,Zhang2014PRA},
where the scattering cross-sections of partial waves with $\ell>0$
are universally determined by the dipole length. The physical explanation
is that scattering at low energy can only occur at distances larger
than the centrifugal barrier, where the potential is dominated by
the dipole-dipole interaction. In addition, the $1/r^{3}$ behavior
of dipole-dipole interaction is weak at large distances, which allows
the application of the perturbative first-order Born approximation.
We apply the first-order Born approximation within the multi-channel
framework, where the $K$-matrix can be approximated by \cite{JiaWang2010PRA,JiaWang2011PRA},
\begin{equation}
\mathcal{K}_{\tau'\tau}^{({\rm Born})}=\pi\int\sum_{\nu'\nu}f_{\nu'\tau'}^{*}(r)2d^{2}\frac{\tilde{B}_{\nu'\nu}^{(3)}}{r^{3}}f_{\nu\tau}(r)dr,
\end{equation}
where $\tilde{B}_{\nu'\nu}^{(3)}=B_{\nu'\nu}^{(3)}/4D$ by comparing
with Eq. (\ref{eq:Vdmatelement}). Inserting the expression of the
regular solution $f_{\nu\tau}$, we arrive at,
\begin{equation}
\mathcal{K}_{\tau'\tau}^{({\rm Born})}=4D\frac{\hbar^{2}\pi}{2\mu}\frac{k_{\tau'}k_{\tau}}{\sqrt{N_{\tau'}N_{\tau}}}\sum_{\nu'\nu}u_{\nu'\tau'}^{*}\tilde{B}_{\nu'\nu}^{(3)}u_{\nu\tau}\Gamma_{\ell'\ell}^{\tau'\tau},\label{eq:KmatBorn}
\end{equation}
where $\Gamma_{\ell'\ell}^{\tau'\tau}=\int drj_{\ell'}(k_{\tau'}r)j_{\ell}(k_{\tau}r)/r$.
Since we are in the perturbative regime, i.e., $\mathcal{K}_{\tau'\tau}^{({\rm Born})}\ll1$,
the cross-section can be approximated by
\begin{equation}
\sigma_{\tau'\tau}^{({\rm Born})}\approx\frac{2\pi}{k_{\tau'}^{2}}|2\mathcal{K}_{\tau'\tau}^{({\rm Born})}|^{2}.\label{eq:crosssectionBorn}
\end{equation}
We would like to remark here that due to the absence of a centrifugal
barrier in the $s$-wave channel, the first-order Born approximation
should not apply to terms in the expansion when $\ell'=\ell=0$ as
the integral $\Gamma_{00}^{\tau'\tau}$ is divergent. However, this
is not a problem, as $\tilde{B}_{\nu'\nu}^{(3)}=0$ for $\ell'=\ell=0$,
and hence gives no contribution to $\mathcal{K}_{\tau'\tau}^{({\rm Born})}$.
Similar to the argument for the non-SOC case \cite{Bohn2009NJP},
the $s$-wave contributions can be included later by supplementing
the Born approximation with a short-range contribution that can be
determined from the full closed-coupling calculations. 

The threshold behavior for partial waves satisfying the first-order
Born approximations can be further explored by using the analytical
properties of the integral $\Gamma_{\ell'\ell}^{\tau'\tau}$:
\begin{multline}
\Gamma_{\ell'\ell}^{\tau'\tau}=\left(\frac{k_{\tau'}}{k_{\tau}}\right)^{\ell'}\frac{\pi\Gamma[\frac{1}{2}(\ell+\ell')]}{8\Gamma[\frac{1}{2}(3+\ell-\ell')]\Gamma(\frac{3}{2}+\ell')}\\
\times{}_{2}F_{1}[\frac{1}{2}(-1-\ell+\ell'),\frac{1}{2}(\ell+\ell'),\frac{3}{2}+\ell',\frac{k_{\tau'}^{2}}{k_{\tau}^{2}}],
\end{multline}
for $k_{\tau}\ge k_{\tau'}$, where $_{2}F_{1}(a,b,c,z)$ is the hypergeometric
function. Due to the symmetry of the integral, one can obtain $\Gamma_{\ell'\ell}^{\tau'\tau}$
for $k_{\tau}<k_{\tau'}$ by simply switching the primed and unprimed
indices on the right-hand side. In addtion, we only need to focus
on the cases with $\ell'-\ell=0,\pm2$, since $\tilde{B}_{\nu'\nu}^{(3)}$
equals zero otherwise. The integral can therefore be further simplified
for $k_{\tau}=k_{\tau'}$,
\begin{equation}
\Gamma_{\ell'\ell}^{\tau'\tau}=\begin{cases}
\frac{1}{2\ell'(\ell'+1)}, & \ell'=\ell,\\
\frac{1}{6(\ell'+1)(\ell'+2)}, & \ell'=\ell-2,\\
\frac{1}{6\ell'(\ell'-1)}, & \ell'=\ell+2.
\end{cases}\label{eq:lambda1}
\end{equation}
For $k_{\tau}\ne k_{\tau'}$, the integral can also be simplified
in the limit $k_{\tau'}/k_{\tau}\ll1$:
\begin{equation}
\Gamma_{\ell'\ell}^{\tau'\tau}\rightarrow\begin{cases}
\frac{(\ell'-1)!2^{\ell'-1}}{(2\ell'+1)!!}\left(\frac{k_{\tau'}}{k_{\tau}}\right)^{\ell'}, & \ell'=\ell,\\
\frac{\ell'!2^{\ell'}}{3(2\ell'+1)!!}\left(\frac{k_{\tau'}}{k_{\tau}}\right)^{\ell'}, & \ell'=\ell-2,\\
\frac{(\ell'-2)!2^{\ell'-2}}{(2\ell'+1)!!}\left(\frac{k_{\tau'}}{k_{\tau}}\right)^{\ell'}, & \ell'=\ell+2.
\end{cases}\label{eq:lambda2}
\end{equation}
where the (double) exclamation marks denote a (double) factorial. 

We remark here that, at the regime $k_{{\rm so}}D\ll kD\ll1$, $k_{\tau}\approx k_{\tau'}\approx k$,
$\Gamma_{\ell'\ell}^{\tau'\tau}$ can therefore always be approximated
by Eq. (\ref{eq:lambda1}). Furthermore, if we investigate the case
of $s_{1}=s_{2}=0$ (and therefore $j=\ell$, $j'=\ell'$ and $m_{j}=m_{j'}=m_{\ell}$),
one can verify that $|2\mathcal{K}_{\tau'\tau}^{({\rm Born})}|\approx|\delta_{\tau'\tau}-\mathcal{S}_{\tau'\tau}^{({\rm Born})}|$
agrees with the $T$-matrix element $|T_{\ell'\ell}^{(m_{\ell}),{\rm Born}}|$
found in Ref. \cite{Bohn2009NJP,Bohn2014PRA} for dipole-dipole scattering
without the presence of SOC.

\section{Example: Two Spin-1/2 Fermionic Dipoles}

We apply our analysis to systems of two identical spin-1/2 fermionic
dipoles as an example, and focus on the $m_{j}=0$ and even $j$ channels.
In order to simplify notification we use $+$ and $-$ to represent
the helicity $+1/2$ and $-1/2$ respectively in this section. Furthermore,
to avoid double counting, we always choose $\xi<\zeta$. Therefore,
the three possible two-particle helicity states for the two dipoles
are given by $(\xi,\zeta)=(-,-),\ (-,+)$ and $(+,+)$, with canonical
momentum $k_{--}=\sqrt{k^{2}+(k_{{\rm so}}/2)^{2}}+k_{{\rm so}}/2$,
$k_{-+}=k$, and $k_{++}=\sqrt{k^{2}+(k_{{\rm so}}/2)^{2}}-k_{{\rm so}}/2$,
and normalization constant $N_{--}=N_{++}=\pi\hbar^{2}\sqrt{k^{2}+(k_{{\rm so}}/2)^{2}}/2\mu$,
and $N_{-+}=\pi\hbar^{2}\sqrt{k}/2\mu$. Using Eq. (\ref{eq:utransformation}),
we find that for $j=0$, only $(-,-)$ and $(+,+)$ are involved and
coupled to the channel functions $\nu=\{j=0,m_{j}=0,\ell=0,s=0\}$
and $\{0,0,1,1\}$. For higher even $j$'s, all three possible helicity
states are involved and coupled to $\nu=\{j,0,j,0\}$, $\{j,0,j-1,1\}$
and $\{j,0,j+1,1\}$. Therefore, only the partial waves with $j=0$
are coupled to $s$-wave, and the first-order Born approximation can
be applied to all other higher partial waves.

\begin{figure}
\includegraphics[width=0.98\columnwidth]{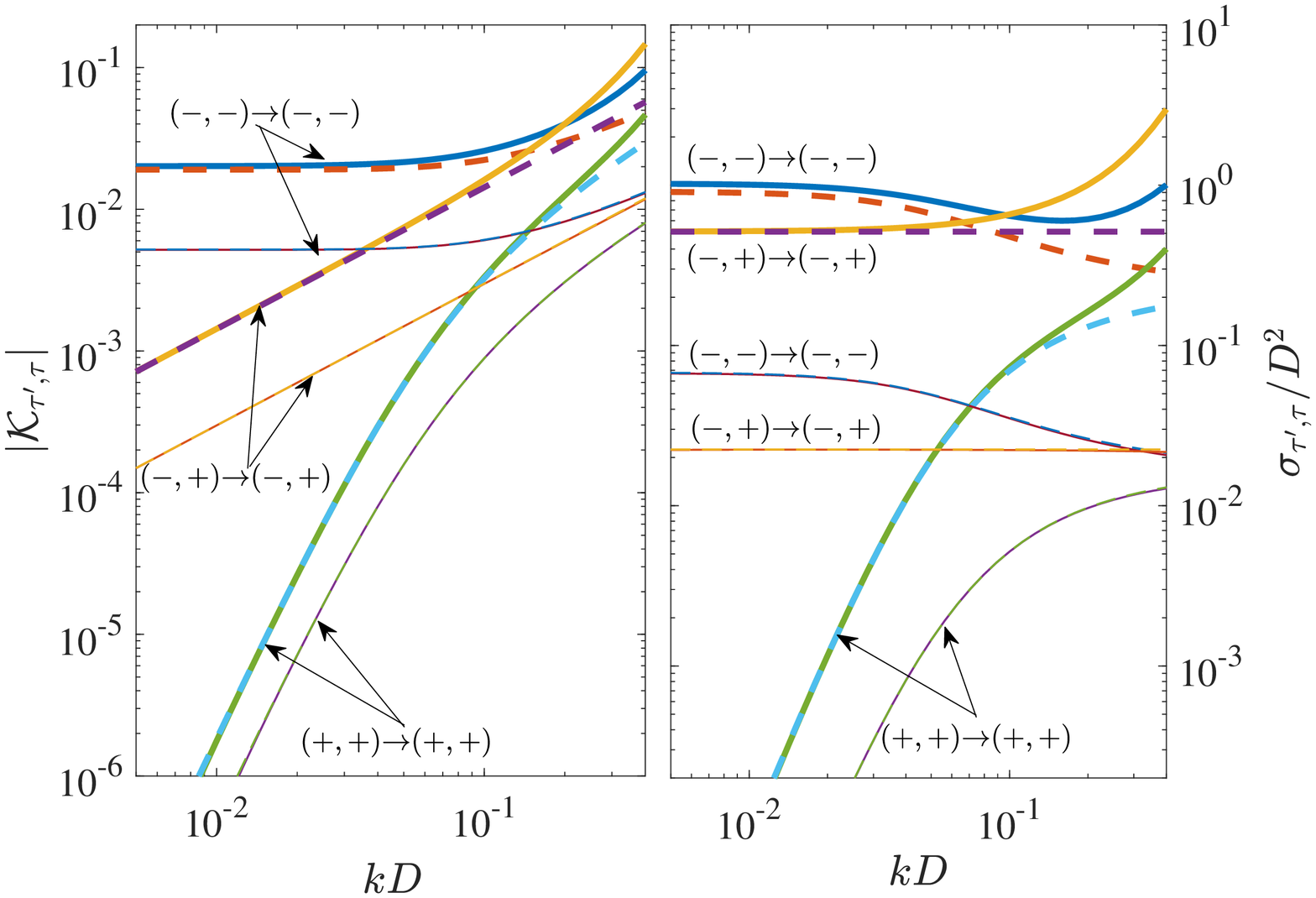}

\caption{(Color online) $K$-matrix elements and cross-sections for elastic
scattering in channels of $(j,m_{j})=(2,0)$, represented by the thick
curves and $(4,0)$, by the thin curves. The incoming and outgoing
helicity states for each curve are indicated on the figure. The solid
curves are results from numerical calculation with $k_{{\rm so}}D=0.1$
and $r_{c}=0.22D$, compared with the dahsed curves from the first-order
Born approximation. For the $(j,m_{j})=(4,0)$ channels, the dashed
curves and solid curves are essentially on top of each other and cannot
be distinguished visually at this scale. \label{fig:elastic}}
\end{figure}

\begin{figure}
\includegraphics[width=0.98\columnwidth]{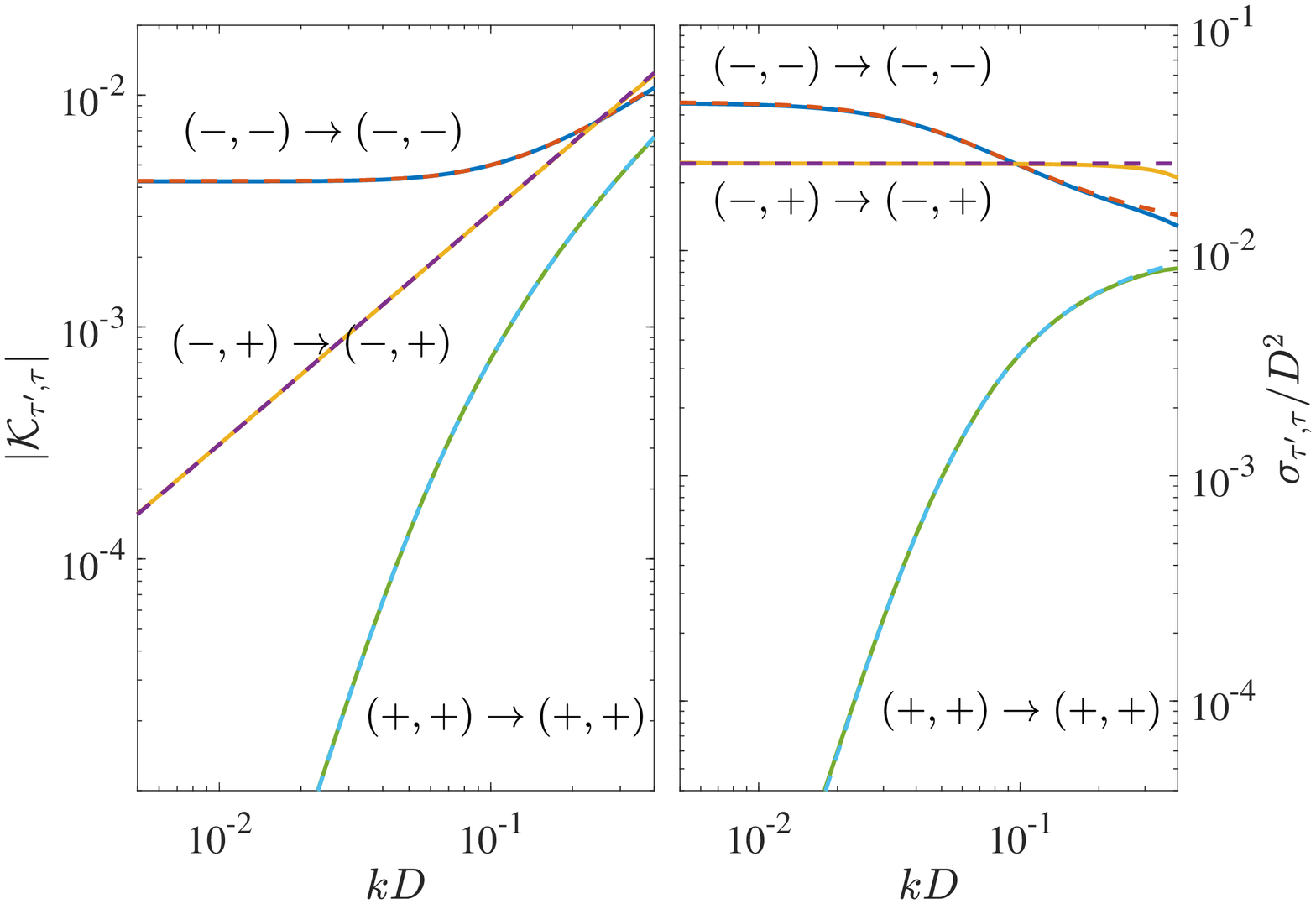}

\caption{(Color online) $K$-matrix elements and cross-sections for inelastic
scattering from channels of $(j,m_{j})=(2,0)$ to $(4,0)$, where
the scattered state remains the same helicity. Other parameters and
notation are the same as Fig. \ref{fig:elastic}.\label{fig:inelastic1}}
\end{figure}
Some of the elastic scattering cross-sections and corresponding $K$-matrix
elements are shown in Fig. \ref{fig:elastic}. The solid curves are
numerical calculations using parameters $k_{{\rm so}}D=0.1$ and $r_{c}=0.22D$,
while the dashed curves are the first-order Born approximations found
from Eq. (\ref{eq:KmatBorn}) and Eq. (\ref{eq:crosssectionBorn}).
The first-order Born approximation agrees excellently with the essentially
exact calculations at low scattering energy (small $k$) and high
angular momentum partial waves. The almost perfect agreement can be
understood by realizing the scattering occurs at larger distances
for lower scattering energy and a higher centrifugal barrier, where
the first-order Born approximation becomes almost exact. 

The first-order Born approximation also agrees well for the essentially
numerically exact inelastic process. In Fig. \ref{fig:inelastic1},
we present some inelastic cross-sections and the corresponding $K$-matrix
elements from channels of $(j,m_{j})=(2,0)$ to $(4,0)$, but the
state remains in the same helicity, which also show very good agreement.
We can write out the explicit form for the elastic and inelastic cross-sections
whose incoming and outgoing two-particle helicity states are the same:
\begin{align}
\sigma_{(--)\rightarrow(--)}^{j,m_{j}\rightarrow j'm_{j}'} & =128\pi D^{2}C_{\tau'\tau}^{2}\frac{\left(\sqrt{k^{2}+(k_{{\rm so}}/2)^{2}}+k_{{\rm so}}/2\right)^{2}}{k^{2}+(k_{{\rm so}}/2)^{2}},\label{eq:BornCrosssectionPartial}\\
\sigma_{(-+)\rightarrow(-+)}^{j,m_{j}\rightarrow j'm_{j}'} & =128\pi D^{2}C_{\tau'\tau}^{2},\\
\sigma_{(++)\rightarrow(++)}^{j,m_{j}\rightarrow j'm_{j}'} & =128\pi D^{2}C_{\tau'\tau}^{2}\frac{\left(\sqrt{k^{2}+(k_{{\rm so}}/2)^{2}}-k_{{\rm so}}/2\right)^{2}}{k^{2}+(k_{{\rm so}}/2)^{2}},
\end{align}
where $C_{\tau'\tau}=\sum_{\nu'\nu}u_{\nu'\tau'}^{*}\tilde{B}_{\nu'\nu}^{(3)}u_{\nu\tau}\Gamma_{\ell'\ell}^{\tau'\tau}$
is a constant with respect to $k$, and $\Gamma_{\ell'\ell}^{\tau'\tau}$
is given by Eq. (\ref{eq:lambda1}). The threshold power law at $k\ll k_{{\rm so}}$
can therefore be given by $\sigma_{(--)\rightarrow(--)}^{j,m_{j}\rightarrow j'm_{j}'}\rightarrow k^{0}$,
$\sigma_{(-+)\rightarrow(-+)}^{j,m_{j}\rightarrow j'm_{j}'}\rightarrow k^{0}$,
and $\sigma_{(++)\rightarrow(++)}^{j,m_{j}\rightarrow j'm_{j}'}\rightarrow k^{4}$.

\begin{figure}
\includegraphics[width=0.98\columnwidth]{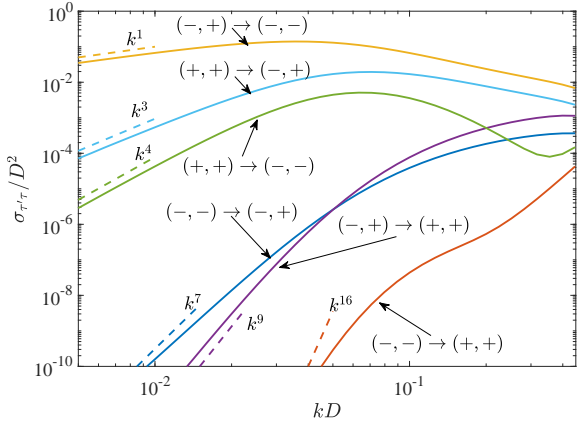}

\caption{(Color online) The cross-sections for inelastic scattering that changes
helicity from channels of $(j,m_{j})=(2,0)$ to $(4,0)$, where the
initial and final helicity are indicated on the figure. The solid
curves are numerical calculations with the same parameters as Fig.
\ref{fig:elastic}. The dashed curves show the corresponding power
law. \label{fig:inelastic2}}
\end{figure}

For inelastic scattering processes with different incoming and outgoing
helicity states, simple formulas for cross-sections without involving
the hypergeometric function do not exist. However, in the limit $k\ll k_{{\rm so}}$,
the threshold behavior can still be analyzed by using Eq. (\ref{eq:lambda2}).
For example, in the process $\tau'=\{2,0,-,+\}\rightarrow\tau=\{4,0,-,-\}$,
we have $k_{-+}/k_{--}\rightarrow k$ in the low $k$ limit, which
leads to $\Gamma_{\ell'\ell}^{\tau'\tau}\rightarrow k^{\ell'}$. Therefore,
the leading order of the $K$-matrix element is $\mathcal{K}_{\tau'\tau}\rightarrow k^{\ell_{{\rm min}}+1/2}$,
where $\ell_{{\rm min}}$ is the lowest $\ell'$ that can couple to
$j'=2$, which equals 1 in this case. The additional factor of $1/2$
in the exponent comes from the factor $k_{\tau'}k_{\tau}/\sqrt{N_{\tau'}N_{\tau}}$.
The cross-section, therefore, obeys the power law $\sigma_{\tau'\tau}\rightarrow k^{1}$.
The same analysis can be applied to other scattering processes, and
are summarized in Fig. \ref{fig:inelastic2} for cross-sections from
channels of $(j,m_{j})=(2,0)$ to $(4,0)$ . We can see that the power-law
describes the threshold behaviors well.

\begin{figure}
\includegraphics[width=0.98\columnwidth]{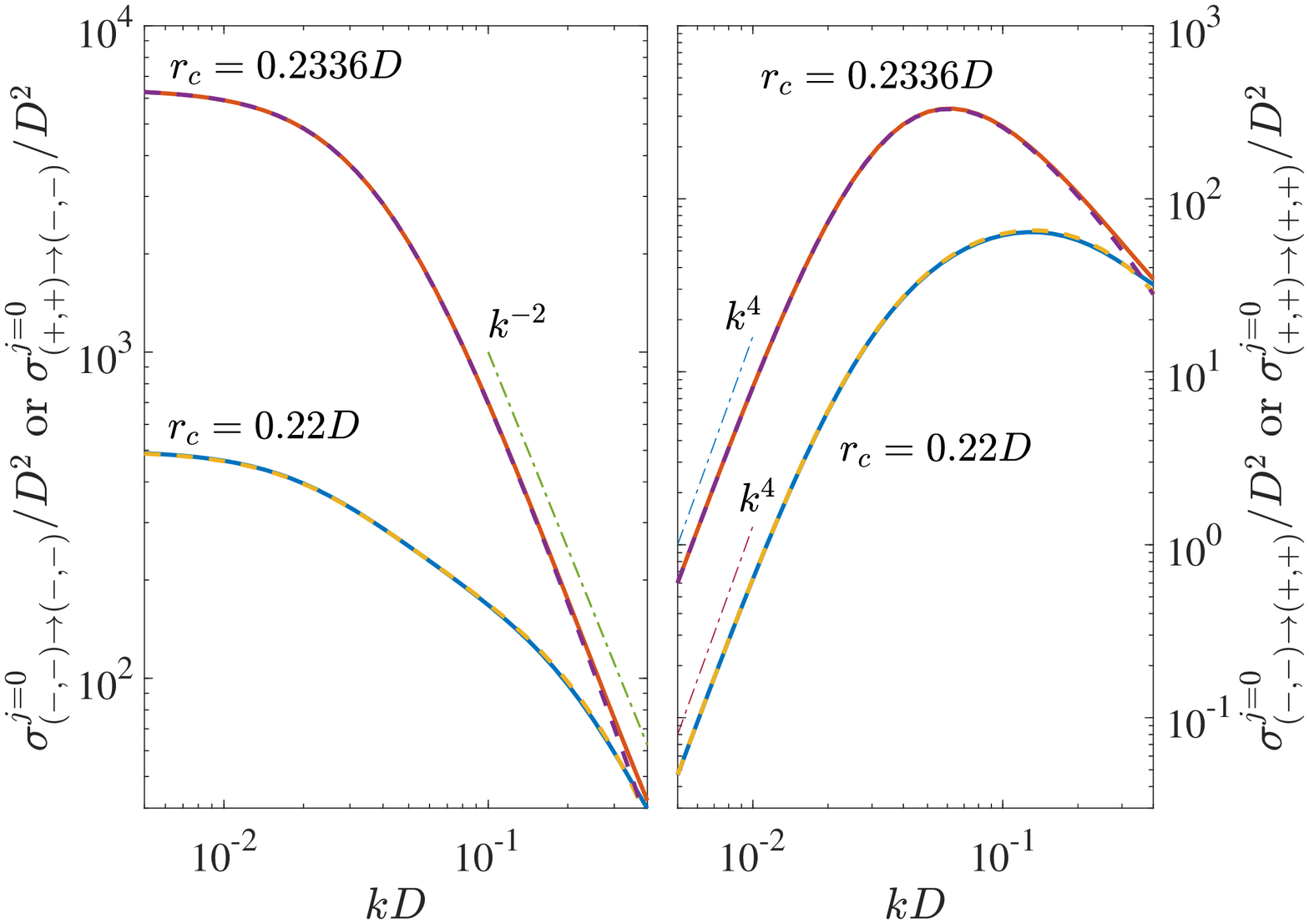}

\caption{(Color online) The cross-sections in the subspace of $(j,m_{j})=(0,0)$.
In the left pannel, the solid/dashed curves shows the cross-sections
from $(-,-)$/$(+,+)$ to $(-,-)$ helicity states, and in the right
pannel, the solid/dashed curves shows the cross-sections from $(+,+)$/$(-,-)$
to $(+,+)$ respectively. These are numerical calculation results
with $k_{{\rm so}}D=0.05$ and different $r_{c}$ indicated on the
figure. For the same set of parameters, the solid and dashed curves
are essentially on top of each other in this scale. \label{fig:sigmaj0}}
\end{figure}

We have also carried out calculations for different $r_{c}$, and
observe that the cross-sections for $j>0$ discussed previously are
insensitive to $r_{c}$, i.e., the cross-sections at low scattering
energy are universally determined by the dipole length $D$ and $k_{{\rm so}}.$
This universality also applies better for lower $k$ and higher angular
momentum $j$ due to the better application of the first-order Born
approximation. However, for the partial cross-sections in the subspace
of $j=0$, the universality implied by the first-order Born approximation
no longer exist due to the absence of a centrifugal barrier. Indeed,
we find that the cross-section in the subspace of $j=0$ depends on
$r_{c}$ and can change by orders of magnitude in our numerical calculation,
as shown in Fig. \ref{fig:sigmaj0}. One can also see that the cross-sections
in the $j=0$ subspace share a lot of similarities with the short-range
results presented in Ref. \cite{Duan2013PRA}, such as the power-law
behaviors, and the identical cross-sections for the same final helicity
state regardless of the initial helicity state at very small $k$.

\begin{figure}
\includegraphics[width=0.98\columnwidth]{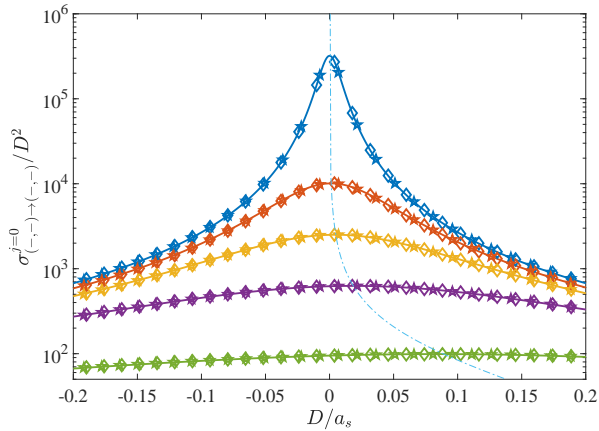}

\caption{(Color online) Elastic cross-sections for the helicity state $(-,-)$
in the subspace of $(j,m_{j})=(0,0)$ for different $k_{{\rm so}}$
as a function of non-SOC singlet $s$-wave scattering length $a_{s}$.
The curves from top to bottom are corresponding to $k_{{\rm so}}D=0.01,\ 0.05,\ 0.1,\ 0.2,$
and $0.5$. The diamod (pentagram) symbols are numerical results calculated
near the resonance at $r_{c}\approx0.232D\ (0.103D)$. The solid curves
are fitted using Eq. (\ref{eq:fitting}). The dash-dotted curve guides
the position of resonances.\label{fig:resonances}}
\end{figure}

In addition, the cross-sections for final states $(-,-)$ at low scattering
energy goes to a constant and can reach resonance by tuning $r_{c}$,
similar to the non-SOC situation studied in Ref. \cite{Kanjilal2008PRA}.
In particular, near the broad potential resonances found in Ref. \cite{Kanjilal2008PRA}
(or more specificly, away from shape resonances and closed-channel
dominated resonances), we have observed another universality, i.e.,
the cross-section can be universally determined by the dipole length
$D$, SOC coupling $k_{{\rm so}}$, and the non-SOC singlet $s$-wave
scattering length $a_{s}$. This universality implies that all the
short-range physics can be absorbed into one single parameter $a_{s}$,
and the detailed form of the short-range interaction is not important.
The underlying physics is the classical suppression of the WKB wave
function amplitude by the large potential well at short distances
\cite{JiaWang2012PRL}, where a frame transformation is allowed. Therefore,
we do not expect this universality can be applied to shape-resonances
or closed-channel dominated resonances, where details of short-range
potential becomes important. 

As shown in Fig. \ref{fig:resonances}, for a fixed $k_{{\rm so}}D$,
the elastic cross-sections $\sigma_{(-,-)\rightarrow(-,-)}^{j=0}$
are calculated near two different resonances and can be fitted as
a function of $a_{s}/D$,

\begin{equation}
\sigma_{(-,-)\rightarrow(-,-)}^{j=0}/D^{2}=\frac{\sigma_{{\rm res}}/D^{2}}{(D/a_{s}-D/a_{{\rm res}})^{2}/\Gamma_{{\rm res}}^{2}+1},\label{eq:fitting}
\end{equation}
where $\sigma_{{\rm res}}$, $a_{{\rm res}}$, and $\Gamma_{{\rm res}}$
are fitting parameters, which only depend on $k_{{\rm so}}D$. Figure
\ref{fig:resonances} also shows the effect of $k_{{\rm so}}$ on
resonances: the resonance shifts further to the positive side and
becomes broader for larger SOC coupling. Interestingly, the shift
of resonance due to the presence of SOC can be explained by the interplay
between the short-range interaction and SOC (see Appendix \ref{sec:Frame-Transformation}
for details).

Another interesting feature in ultracold scattering with the presence
of SOC is that the particles are preferentially scattered into the
lowest helicity states (where the particle\textquoteright s momentum
is antiparallel to its spin direction), regardless of their incidence
channel \cite{Duan2013PRA,WangYujun2012PRA}. This spontaneous handedness
is an analog of an antiferromagnetic phenomena induced by the momentum-dependent
magnetic field. The presence of dipole-dipole interaction would not
change this spontaneous handedness effect, as can be seen by comparing
the ratios of the different scattering cross-sections $\sigma_{\tau'\tau}/\sigma_{\tau\tau'}=k_{\tau}^{2}/k_{\tau'}^{2}$.
Therefore, after sometime, we expect all the particles in our system
to be in a ``$-$'' helicity state, and any rethermalization due
to a perturbation should be described by the total cross-section of
$\sigma_{(-,-)\rightarrow(-,-)}^{{\rm tot}}=\sigma_{(-,-)\rightarrow(-,-)}^{j=0}+\sigma_{(-,-)\rightarrow(-,-)}^{({\rm Born})}$,
where
\begin{equation}
\sigma_{(-,-)\rightarrow(-,-)}^{({\rm Born})}=\sum_{j,j',m_{j}}\sigma_{(-,-)\rightarrow(-,-)}^{j,m_{j}\rightarrow j'm_{j}}.
\end{equation}
Summing the partial cross-section from Eq. (\ref{eq:BornCrosssectionPartial})
in the limit of $k\rightarrow0$ gives $\sigma_{(-,-)\rightarrow(-,-)}^{{\rm tot}}\approx\sigma_{(-,-)\rightarrow(-,-)}^{j=0}+4.46D^{2}$.
Noticing that $(32\pi D^{2}/15+32\pi D^{2}/45)/2\approx4.46D^{2}$
implies that the total cross-section from the first-order Born approximation
equals the average of cross-sections for identical fermions $(32\pi D^{2}/15\approx6.70D^{2})$
and identical bosons $(32\pi D^{2}/45\approx2.23D^{2})$ without the
presence of SOC \cite{Aikawa2014PRL,Bohn2014PRA}. We believe this
reflects the fact that the total cross-section sums over the singlet
and triplet cross-sections, which corresponds to the non-spin identical
bosons and fermions respectively. In addition, when the particles
are in $(-,-)$ helicity, they have equal probability to be projected
into singlet and triplet states. 

\section{Conclusion}

In summary, this paper extends previous theoretical studies of ultracold
scattering in the presence of 3D isotropic SOC to a dipolar system.
Our formalism is general in the sense that it can be applied to either
bosons or fermions with arbitrary spin and the inclusion of any angular
momentum partial waves. Similar to the non-SOC cases, the cross-sections
involving high angular momentum partial waves can be well described
by the first-order Born approximation, and can be determined universally
by the dipole length $D$ and spin-orbit coupling strength $k_{{\rm so}}.$
However, the cross-sections that can couple to an $s$-wave channel
depend on the short-range physics and can have resonances. Nevertheless,
all the short-range physics can be described by one additional parameter,
$a_{s}$, near a broad potential resonance. We have tested our theory
in the example system of spin-$1/2$ dipolar fermions, and find excellent
agreement with our numerical calculations. 

While this work focuses on the ultracold regime $E\rightarrow0^{+}$,
our formalism can be easily extended to the negative scattering energy
$E=-\kappa^{2}/2\mu$ following the same approach in Ref. \cite{Guan2016PRA}.
In this energy regime, however, the canonical momentum should be understood
as given by $k_{\xi\zeta}=\sqrt{\kappa_{\xi\zeta}^{2}-\kappa^{2}}-\kappa_{\xi\zeta}$
(which has the same definition of $\bar{k}$ in Ref. \cite{Guan2016PRA}).
This topic, however, will be addressed elsewhere in the future.
\begin{acknowledgments}
We would like to thank Hui Hu for his insightful discussions. This
research was supported under Australian Research Council\textquoteright s
Future Fellowships funding scheme (project number FT140100003) and
Discovery Projects funding scheme (project number DP170104008). The
numerical calculations were partly performed using Swinburne new high-performance
computing resources (Green II).
\end{acknowledgments}

\appendix

\section{Propagation Method\label{sec:Propagation-Method}}

Based on the R-matrix propagation method using discrete variable representations
(DVR) basis, we develop a numerically stable method for propagating
the logarithmic derivative matrix $\underline{\mathcal{L}}=\underline{F}'\underline{F}^{-1}$,
where $\underline{F}$ is the solution of Eq. (\ref{eq:radialequation}).
Numerically, we seperate the whole regime into many sectors. For each
sector $r\in[a_{1},a_{2}]$ the propagation method allows us to calculate
the logarithmic derivative matrix at one end $\underline{\mathcal{L}}(a_{2})$
for a given $\underline{\mathcal{L}}(a_{1})$ on the other end. One
key ingredient in this method is the construction of the DVR basis.
Our DVR basis functions $\pi_{j}(r)$ are in the form of Langrange
polynomial:

\begin{equation}
\pi_{i}(r)=\sqrt{\frac{1}{\tilde{w_{i}}}}\prod_{j\neq i}^{N}\frac{r-r_{j}}{r_{i}-r_{j}},
\end{equation}
where $r_{i}=sx_{i}+\bar{a}$ and $\tilde{w}_{i}=sw_{i}$ are defined
by the $N$ Gauss-Lobatto quadrature points $x_{i}$ and weights $w_{i}$
correspondingly, with $s=(a_{2}-a_{1})/2$ and $\bar{a}=(a_{2}+a_{1})/2$
\cite{Rescigno2000PRA}. One can easily verify that a DVR basis satisfies
$\pi_{i}(r_{j})=\delta_{ij}/\sqrt{\tilde{w_{i}}}$ that leads to the
DVR approximation, i.e. $\intop_{a_{1}}^{a_{2}}\pi_{i}(r)v(r)\pi_{j}(r)dr\approx v(r_{i})\delta_{ij}$
with a smooth function $v(r)$. In addition, the derivative of DVR
basis $\pi_{i}'(r)$ can be derived analytically. 

Under the DVR approximation, Eq. (\ref{eq:radialequation}) can be
written as,

\begin{equation}
\underline{H}\vec{c}^{(\mu)}\equiv(\underline{T}+\underline{M}+\underline{V}-k^{2}\underline{I})\vec{c}^{(\mu)}=\underline{\Lambda}\vec{c}^{(\mu)},\label{eq:matrixequation}
\end{equation}
after integrating by parts, where $c_{\nu j}^{(\mu)}$ (in vector
notation, $\vec{c}^{(\mu)}$) are the expansion coefficients for the
matrix elements of $\underline{F}$,

\begin{equation}
F_{\nu\mu}(r)=\sum_{j}c_{\nu j}^{(\mu)}\pi_{j}(r).
\end{equation}
 The matrix elements of other terms in Eq. (\ref{eq:matrixequation})
are given by

\begin{equation}
T_{\mu i,\nu j}=\delta_{\mu\nu}\sum_{m}\tilde{w}_{m}\pi_{i}'(r_{m})\pi_{j}'(r_{m}),
\end{equation}

\begin{equation}
M_{\mu i,\nu j}=A_{\mu\nu}\frac{1}{2}\sum_{m}\tilde{w}_{m}\left[\pi_{i}(r_{m})\pi_{j}'(r_{m})-\pi_{i}'(r_{m})\pi_{j}(r_{m})\right],
\end{equation}
\begin{equation}
V_{\mu i,\nu j}=B_{\mu\nu}(r_{i})\delta_{ij},
\end{equation}
 which are all symmetric. The surface term is given by

\begin{equation}
\Lambda_{\mu i,\nu j}=\{\pi_{i}(r)[\delta_{\mu\nu}\pi_{j}'(r)-\frac{1}{2}A_{\mu\nu}\pi_{j}(r)]\}|_{a_{1}}^{a_{2}}.
\end{equation}
From the form of the surface term, we define a matrix $\underline{L}=\mathcal{\underline{L}}-\underline{A}/2$,
so that we have,

\begin{equation}
\underline{F}'(r)=\left[\underline{L}(r)+\frac{1}{2}\underline{A}\right]\underline{F}(r),
\end{equation}
which gives,
\begin{equation}
\sum_{j\nu}\Lambda_{i\tau,j\nu}c_{j\nu}^{(\mu)}=\sum_{\nu}\left[\frac{\delta_{iN}\delta_{jN}}{\tilde{w}_{N}}L_{\mu\nu}(a_{2})c_{\nu N}^{(\mu)}-\frac{\delta_{i1}\delta_{j1}}{\tilde{w}_{1}}L_{\mu\nu}(a_{1})c_{\nu1}^{(\mu)}\right].
\end{equation}
Defining the matrix $\underline{h^{cc'}}$ with the elements
\begin{equation}
h_{i\tau,j\nu}^{cc'}=\underline{H}{}_{i\tau,j\nu}+\frac{\delta_{i1}\delta_{j1}}{\tilde{w}_{1}}L_{\mu\nu}(a_{1}),
\end{equation}
where $c$ and $c'$ are a collective index for selected DVR basis
indices, i.e. $i\in c$ and $j\in c'$. The radial equation can therefore
be written in a matrix form:
\begin{equation}
\begin{pmatrix}h^{ss} & h^{sN}\\
h^{Ns} & h^{NN}
\end{pmatrix}\begin{pmatrix}\vec{c}_{s}^{(\mu)}\\
\vec{c}_{N}^{(\mu)}
\end{pmatrix}=\begin{pmatrix}0 & 0\\
0 & L(a_{2})/\tilde{w}_{N}
\end{pmatrix}\begin{pmatrix}\vec{c}_{s}^{(\mu)}\\
\vec{c}_{N}^{(\mu)}
\end{pmatrix},
\end{equation}
which leads to

\begin{equation}
\frac{1}{\tilde{w}_{N}}\underline{L}(a_{2})=\underline{h^{NN}}-\underline{h^{Ns}}\frac{1}{\underline{h^{ss}}}\underline{h^{sN}},
\end{equation}
with $s=1,2,...,N-1$. For the first sector, if we impose a hard-wall
boundary condition from the left, i.e. $F(a_{1})=0$, a specitial
treatment has to be implemented for this sector. Notice that only
the first DVR basis $\pi_{1}(r)$ has non-zero value at $a_{1},$
the boundary condition can be easily satisfied by chossing $s=2,...,N-1$
for the first sector. An additional feature from this formalism is
that one can easily see that $\underline{L}(a_{2})$ is automaticlly
real and symmetric if $\underline{L}(a_{1})$ is real and symmetric,
implying $\underline{L}$ is real and symmetric everywhere if we impose
the hard-wall boundary condition.

\section{Symmetry of $K$-matrix\label{sec:Symmetry-of--matrix}}

As we will show later, the $K$-matrix $\underline{\mathcal{K}}$
is real and symmetric (correspondingly, the S-matrix $\underline{\mathcal{S}}$
is a unitary matrix) as long as $\underline{L}$ is real and symmetric.
We have already shown that $\underline{L}$ is real and symmetric
in the DVR formulation. Below, we give a more general proof without
the help of any specific radial basis. From the definition of $\underline{L}$,
we have,

\begin{equation}
F'-\frac{1}{2}AF=LF.
\end{equation}
From hereon in this section, we neglect the underline for matrix variables
to sympify the notation. From the radial equation Eq. (\ref{eq:radialequation}),
we have,
\begin{equation}
F''=AF+B-k^{2}I.
\end{equation}
After some algebra, the derivative of $L$ is given by,
\begin{equation}
L'=B-k^{2}I+\left(\frac{1}{2}A-L\right)F'F^{-1}.
\end{equation}
The definition of $L$ can also be re-written as
\begin{equation}
F'F^{-1}=\frac{1}{2}A+L.
\end{equation}
Finally, we have,
\begin{equation}
L'=B-k^{2}I+\left(\frac{1}{2}A-L\right)\left(\frac{1}{2}A+L\right).
\end{equation}
Notice that $B$ is real and symmetric, and $A$ is real and antisymmetric.
Therefore, if $L$ is real and symmetric, $L'$ must also be real
and symmetric, which implies $L$ is real and symmetric at all points
as long as it is real and symmetric at one point, which is usually
satisfied at the origin. 

In order to show that $\underline{\mathcal{K}}$ is also real and
symmetric, we need to apply the properties of Wronskian of the regular
and irregular solutions givn by \cite{WangSuJu2015PRA},
\begin{equation}
f'^{T}f-f^{T}f'+f^{T}Af=0,
\end{equation}
\begin{equation}
g'^{T}g-g^{T}g'+g^{T}Ag=0,
\end{equation}
and
\begin{equation}
f'^{T}g-f^{T}g'+f^{T}Ag=\frac{1}{\pi}I.
\end{equation}
The radial wave function can be expressed in terms of these regular
and irregular solutions as $F=f-g\mathcal{K}$. Since $f$ and $g$
are both real, $\mathcal{K}$ is automatically guaranteed to be real.
Substituting the regular and iregular solution into the definition
of $L$ gives,
\begin{equation}
(f'-g'\mathcal{K})-\frac{1}{2}A(f-g\mathcal{K})=L(f-g\mathcal{K}),
\end{equation}
which can be rewritten as,
\begin{equation}
L^{-1}\left[(f'-g'\mathcal{K})-\frac{1}{2}A(f-g\mathcal{K})\right]=(f-g\mathcal{K}),
\end{equation}
 and
\begin{equation}
(f^{T}-\mathcal{K}^{T}g^{T})L=(f'^{T}-\mathcal{K}^{T}g'^{T})+\frac{1}{2}(f^{T}-\mathcal{K}^{T}g^{T})A.
\end{equation}
 Multiplying both the left and right-hand-side of these two equations
gives,
\begin{multline}
(f^{T}-\mathcal{K}^{T}g^{T})\left[(f'-g'\mathcal{K})-\frac{1}{2}A(f-g\mathcal{K})\right]=\\
\left[(f'^{T}-\mathcal{K}^{T}g'^{T})+\frac{1}{2}(f^{T}-\mathcal{K}^{T}g^{T})A\right](f-g\mathcal{K}).
\end{multline}
 With the help of Wronskian, the above equation gives $\mathcal{K}=\mathcal{K}^{T}$,
poving that $\mathcal{K}$ is symmetric. 

Since $\mathcal{K}$ is real and symmetric, there exists a unitary
transofrmation $U$ that can diagonlize $\mathcal{K}$: $U^{\dagger}\mathcal{K}U=\kappa$,
where $\kappa$ is a diagonalized and real matrix. This unitrary transformation
$U$ can also diagonalize $I+i\mathcal{K}$ and $I-i\mathcal{K}$,
which implies,

\begin{equation}
\mathcal{S}\equiv(I+i\mathcal{K})(I-i\mathcal{K})^{-1}=(I-i\mathcal{K})^{-1}(I+i\mathcal{K}),
\end{equation}
 is a unitary matrix.

\section{Frame Transformation\label{sec:Frame-Transformation}}

\begin{figure}
\includegraphics[width=0.98\columnwidth]{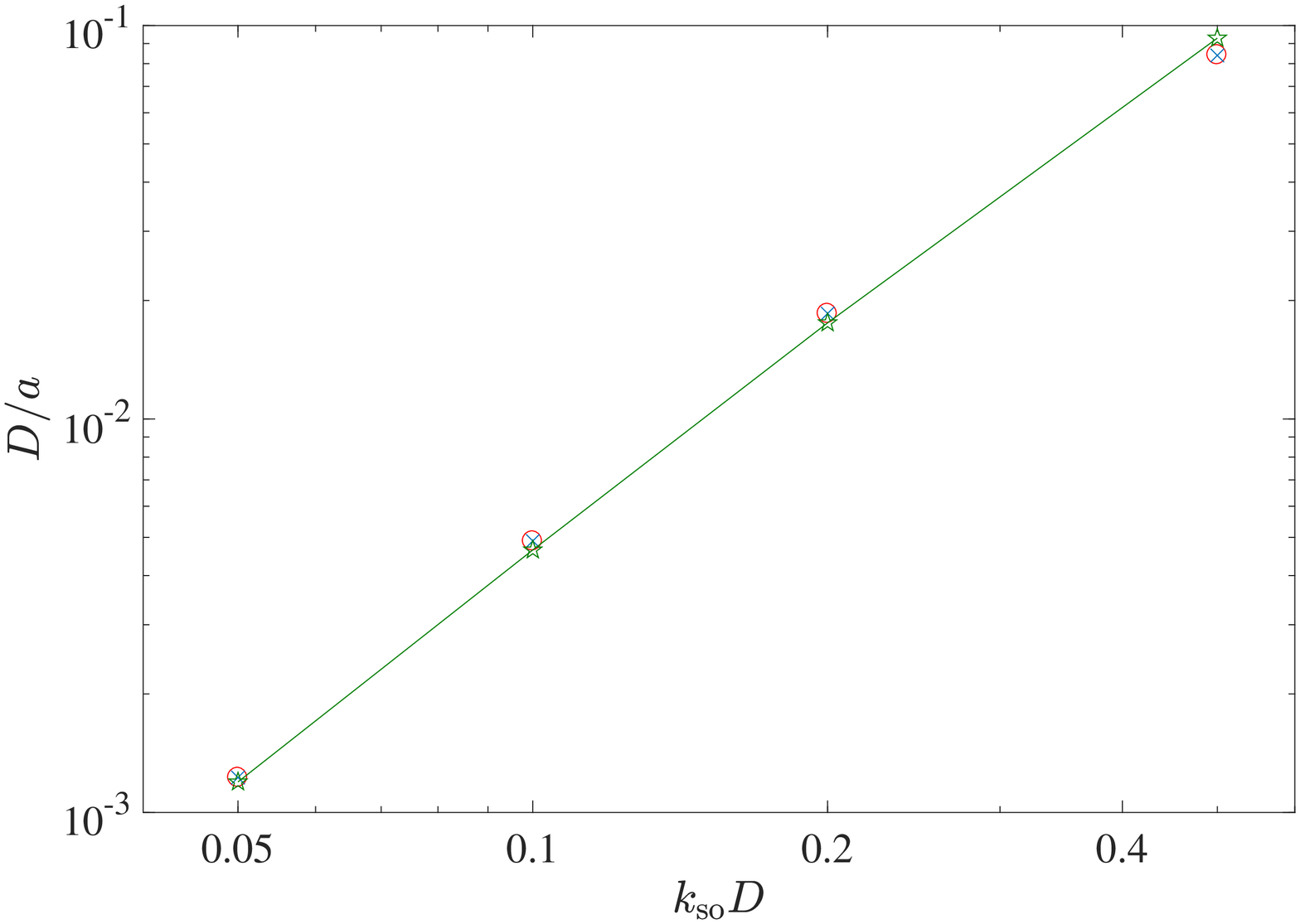}

\caption{(Color online) Comparison of $D/a_{s}^{*}$ (shown by green pentagrams)
and $D/a_{{\rm res}}$ {[}shown by red circles (blue crosses) for
resonance near $r_{c}\approx0.232D$ ($r_{c}\approx0.103D$){]} as
a function of SOC strength $k_{{\rm so}}D$. \label{fig:ares}}
\end{figure}

One elegant discovery in Ref. \cite{Guan2017PRA} is that, a frame
transformation can be found if the atom-atom interaction is short-range.
Here, we extend their approach from the distinguishable particle case
to the identical particle case. Under such a transformation, the short-range
hamiltonian $H^{{\rm SR}}$ and the free-space hamiltonian $H^{{\rm FS}}$
are analytically related by,
\begin{equation}
H^{{\rm SR}}=H^{FS}+\frac{E_{r}}{2\hbar^{2}}[(\vec{s}_{1}-\vec{s}_{2})\cdot\vec{r},(\vec{s}_{1}-\vec{s}_{2})\cdot\vec{\nabla}],\label{eq:frametransformation}
\end{equation}
where $E_{r}=\hbar k_{{\rm so}}^{2}/2m$. Noticing that our definition
of $k_{{\rm so}}$ has a factor of two different than the one defined
in Ref. \cite{Guan2017PRA}. For spin-1/2 atoms, $\vec{s}_{n}=\hbar\vec{\sigma}^{(n)}/2$,
where $\vec{\sigma}^{(n)}$ is a vector whose components are the three
Pauli matrices for the particle $n$. For identical particles, the
second term of the right-hand-side of Eq. (\ref{eq:frametransformation})
can be simplified as,
\begin{equation}
\varepsilon=-\frac{E_{r}}{2\hbar^{2}}\left[\vec{l}\cdot\vec{s}+\left(\vec{s}_{1}-\vec{s}_{2}\right)^{2}\right],
\end{equation}
where $\vec{s}=\vec{s}_{1}+\vec{s}_{2}$ is the total spin operator.
Using tensor spherical harmonics to expand this operator gives a diagonal
matrix for two spin-1/2 particles
\begin{equation}
\varepsilon_{\nu}=-\frac{E_{r}}{4}\left[j(j+1)-\ell(\ell+1)-s(s+1)+6\right],
\end{equation}
where $\nu=\{j,m_{j},\ell,s\}$ are the quantum numbers for the tensor
spherical harmonics. In the spirit of frame transformation \cite{Guan2017PRA},
this effectively replaces the short-range parameter (scattering phase
shift) $\delta_{\nu}(k)$ by $\delta_{\nu}(k_{\nu})$, where $\hbar^{2}k_{\nu}^{2}/2\mu=E-\varepsilon_{\nu}=\hbar^{2}k^{2}/2\mu-\varepsilon_{\nu}$.
Near a $s$-wave resonance in free-space, the $K$-matrix reaches
resonance if $1/a_{s}(k_{s})\rightarrow0$, where $\hbar^{2}k_{s}^{2}/2\mu=\hbar^{2}k^{2}/2\mu+3E_{r}/2$. 

In the presence of long-range and anisotropic dipole-dipole interaction,
a direct application of the approach in Ref. \cite{Guan2017PRA} is
not appropriate. However, we find that the resonance shift near an
$s$-wave resonance described in Fig. \ref{fig:resonances} can still
be understood via this approach, since the interplay between the short-range
interaction and SOC plays an important role here. Near the resonance
of $r_{c}\approx0.232D$ without the presence of SOC, we calculate
the value of $s$-wave scattering length $a_{s}^{*}=a_{s}(0)$ if
the corresponding potential parameters gives $1/a_{s}(k_{s})\rightarrow0$.
Interestingly, as illustrated in Fig. \ref{fig:ares}, the values
of $a_{s}^{*}$ agrees well with $a_{{\rm res}}$ in Eq. (\ref{eq:fitting})
that is calculated by fitting the resonance of a numerical calculation
with the presence of SOC, clearly showing the shift of the resonance
position is a result of interplay between a short-range interaction
and SOC.

\bibliographystyle{apsrev4-1}
\bibliography{Refs}

\begin{thebibliography}{67}
\bibitem [{con()}]{convergence}%
  \BibitemOpen
  \href@noop {} {}\bibinfo {note} {With the same $j_{\rm max}$, we have also
  reproduced the results of Ref. \cite{Kanjilal2008PRA} for dipole-dipole
  scattering without the presence of SOC.}\BibitemShut {Stop}%
\end{thebibliography}

\end{document}